\documentclass[runningheads]{llncs}
\usepackage[T1]{fontenc}
\usepackage{diagbox}
\usepackage[linesnumbered,ruled,vlined]{algorithm2e}
\usepackage{amsmath,amssymb,amsfonts}
\usepackage{xcolor}
\usepackage{color}
\usepackage{makecell} 
\usepackage{multirow} 
\usepackage{booktabs}
\usepackage{threeparttable} 
\usepackage{graphicx}

\begin{document}
\title{DWFS-Obfuscation: Dynamic Weighted Feature Selection for Robust Malware Familial Classification under Obfuscation}
%
\titlerunning{Abbreviated paper title}
If the paper title is too long for the running head, you can set
an abbreviated paper title here
\author{Xingyuan Wei \inst{1,2}\orcidID{0009-0001-6595-4222} \and
Zijun Cheng \inst{4} \and
Ning Li     \inst{1} \and 
Qiujian Lv  \inst{1} \and 
Ziyang Yu \(*\) \inst{1,2} \and 
Degang Sun  \inst{2,3}}
\authorrunning{F. Author et al.}
%
\institute{Institute of Information Engineering, Chinese Academy of Sciences, Beijing, China \email{talentedyuan@gmail.com} \and
University of Chinese Academy of Sciences \and 
Computer Network Information Center Chinese Academy of Sciences Beijing, China \and 
Space Engineering University 
\\ }
\maketitle              
\begin{abstract}

Due to its open-source nature, the Android operating system has consistently been a primary target for attackers. Learning-based methods have made significant progress in the field of Android malware detection. However, traditional detection methods based on static features struggle to identify obfuscated malicious code, while methods relying on dynamic analysis suffer from low efficiency. To address this, we propose a dynamic weighted feature selection method that analyzes the importance and stability of features, calculates scores to filter out the most robust features, and combines these selected features with the program’s structural information. We then utilize graph neural networks for classification, thereby improving the robustness and accuracy of the detection system. We analyzed 8,664 malware samples from eight malware families and tested a total of 44,940 malware variants generated using seven obfuscation strategies. Experiments demonstrate that our proposed method achieves an F1-score of 95.56\% on the unobfuscated dataset and 92.28\% on the obfuscated dataset, indicating that the model can effectively detect obfuscated malware.

\end{abstract}

\section{Introduction}
Malware, particularly on the Android mobile platform, poses an increasingly severe threat. According to a report by AVTEST, as of January 2025, the number of Android malware instances reached 35,641,466\cite{AVTEST}, resulting in substantial losses. Due to the Android platform’s ease of cloning software, many vendors and malware authors employ obfuscation strategies to protect software intellectual property. However, Malware creators also frequently use obfuscation techniques to conceal malicious code, making malware detection significantly more challenging.

To detect Android malware, existing research has utilized machine learning methods\cite{drebin},\cite{naticusdroid} and deep learning approaches\cite{msdroid},\cite{malscan},\cite{ast} to analyze code. Current static analysis methods, such as those extracting opcodes, API calls, and permission requests or relying on single feature types, offer high execution efficiency and accuracy. However, their performance degrades significantly under the interference of code obfuscation strategies. Dynamic analysis methods\cite{liceNovel},\cite{dynamicSurvey}, which require running the software to obtain features, can enhance detection robustness but suffer from low efficiency. The feature subsets of Android applications are virtually inexhaustible, and inputting all features into a model results in low efficiency and excessive redundant data. Therefore, identifying a subset of features from a vast feature pool that can resist the effects of obfuscation techniques is particularly critical.

Although obfuscation techniques can, to some extent, alter the code structure of an application and cause changes at the code level, such as in information flow, the core operational logic of the application should still exhibit significant similarity. We filter out features that are minimally affected by obfuscation, which we refer to as Anti-Obfuscation features. To this end, we propose the Dynamic Weighted Feature Selection (DWFS) algorithm, integrating the selected features with Graph Neural Networks (GNNs). Specifically: 1) Extract a rich candidate feature set from a large number of Android APK files. 2) Filter out features from this vast candidate set that are both significant for malware detection and resistant to obfuscation. 3) Construct a method-level Sensitive Behavior Subgraph (SBS) by parsing APK files to capture the program’s behavioral information. Assign the features selected in step 1) to each node in the SBS, generating feature vectors. Finally, leverage GNNs to learn a joint representation of the SBS’s graph structure and node features, enabling malware family classification.



 Our main contributions are as follows:
 
 (1) Malware Familial Classification. We design a novel framework for malware familial classification that integrates obfuscation-resistant features and the program’s structural information, utilizing GNNs for classification.

 (2) Anti-Obfuscation. We innovatively propose a dynamic weighted feature selection method that analyzes the importance and stability of features to automatically select the most robust ones. Additionally, we make these features publicly available, providing empirical insights for future research on the robustness of malware detection. We also open-source our code, which can be accessed from GitHub \footnote{https://github.com/XingYuanWei/DWFS-Obfuscation}.

 (3) Effective Detection. We optimize the sensitive behavior subgraph derivation algorithm, combining the selected obfuscation-resistant factors with GNNs to efficiently capture the malicious behaviors of programs, thereby enhancing the accuracy and robustness of the malware detection system.

\section{Related Work}
\subsection{Android Malware Familial Classification Based On Static Analysis}
Previous research works\cite{constrasive},\cite{falDroid},\cite{apaposcopy},\cite{mudFlow} have proposed dividing malware samples into their respective families. Malware family classification based on static analysis has achieved promising results. For example, FalDroid\cite{falDroid} conducts frequent subgraph analysis to extract common subgraphs for each family and uses them to perform familial classification. Apaposcopy\cite{apaposcopy} considers both data flow and control flow information from malware samples, classifying them by performing weighted program analysis. These methods leverage different types of program information to achieve accurate malware classification; however, none of them account for the impact of obfuscation on family classification.

\subsection{Anti-obfuscation Android Malware Familial Classification}

AndrODet\cite{andrODet} is one of the Android obfuscation detectors. This system detects obfuscated applications using online machine learning algorithms, with features statically extracted from bytecode. It achieves an accuracy of 92.02\% in string encryption detection, 81.41\% in identifying string encryption, and 68.32\% in control flow obfuscation detection. Orlis\cite{orlis} is a library detector for Android applications. It extracts features from function calls and Android API calls, making them resilient to the most common obfuscation techniques. RevealDroid\cite{revealDroid} is an obfuscation-resilient malware classifier. The features used to train the model are derived from Android API usage, reflection, and application permissions. While it achieves strong results (98\% accuracy in detecting malware and 95\% accuracy in identifying its family), the obfuscation techniques considered in its experiments are relatively simple.

\subsection{Dynamic Weighted Feature Selection}
Feature selection contributes to improving machine learning performance by selecting a subset of relevant features for the learning algorithm. The DFWS method proposed in this paper resembles an adversarial defense approach, aiming to enhance model performance by selecting robust features, particularly when confronting code obfuscation in Android malware families. Our method evaluates the stability and importance of features under malware code obfuscation, selecting those that remain reliable even after obfuscation. The earliest similar idea was proposed by Sun et al.\cite{sunDynamicWeight}, who introduced a dynamic weighted feature selection method based on feature interactions, demonstrating its effectiveness on four gene microarray datasets. Meanwhile, we draw inspiration from the robustness of feature selection in adversarial machine learning\cite{featureSelectionAdversarilly}, combining dynamic weighting with adversarial machine learning principles. A related work, DroidRL\cite{droidRL}, leverages a Reinforcement Learning algorithm to obtain a feature subset effective for malware classification. This framework’s feature selection exhibits high performance without any manual intervention, but its feature selection process remains static.

\section{Methodology}
\begin{figure}[!t]
\centering
\includegraphics[scale=0.43]{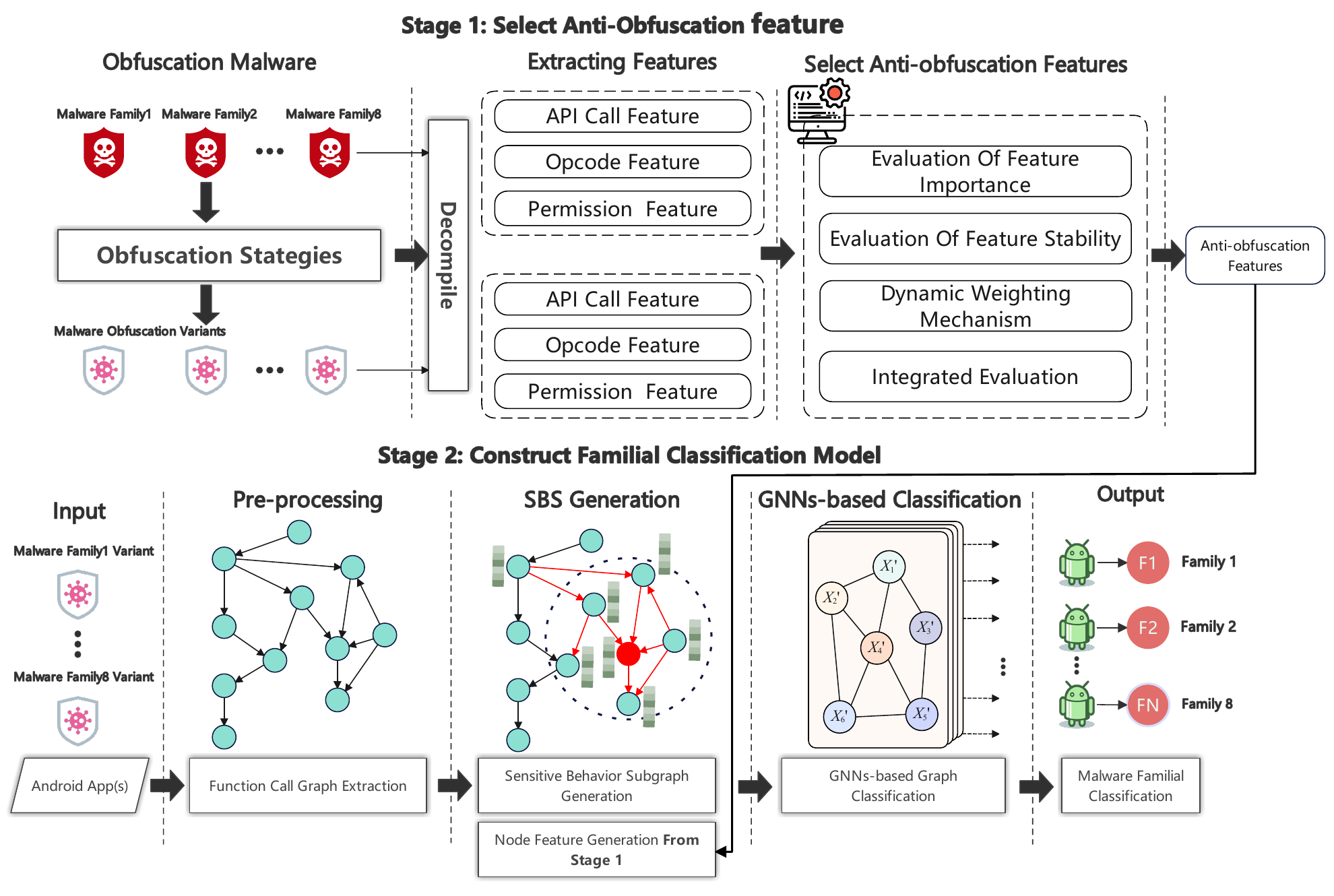}
\caption{DWFS-Obfuscation Overview.}
\label{fig_overview}
\end{figure}

This paper proposes a malware detection method for Android APK files that combines a dynamic weighted feature selection algorithm with graph neural networks. The overall architecture of the system is illustrated in Figure \ref{fig_overview}, consisting of two main stages: Stage 1, including 1) obfuscating malware, 2) feature extraction, and 3) dynamic weighted feature selection; Stage 2, including 1) preprocessing to extract the Function Call Graph (FCG), 2) constructing the Sensitive Behavior Subgraph and generating node features, and 3) GNNs training and classification.

\subsection{Stage 1}\label{subsection:stage1}
\subsubsection{Feature Extraction}




Feature extraction is a foundational step in malware detection. This paper extracts the following three types of static features from Android APK files, forming a high-dimensional candidate feature set:

Opcodes, By analyzing the Dalvik bytecode of the APK, we extracted all Dalvik opcodes from the official Android website\cite{androidWebsite}. The frequency of opcodes can reflect the behavior of a program and its execution logic. For example, opcodes such as move and invoke-virtual have a total of 235.

API Calls, We utilized APIChecker\cite{apichecker} to statistic the usage of Android APIs. We focus on analyzing sensitive APIs, including sendTextMessage, getDeviceId, and others have a total of 426, which reflect the functional behavior of the program.

Permissions, We record the permissions requested by the APK, represented as binary features, such as READ\_SMS and INTERNET, which indicate the program’s access requirements to system resources. We include a total of 86 commonly used Android permission features in our overall set, sourced from NATICUSdroid\cite{naticusdroid}.

Our goal is to select Anti-Obfuscation features from the 747 feature dimensions mentioned above, reducing the dimensionality of the features. In this process, we achieve a refinement of the information, thereby increasing its density.

\subsubsection{Dynamic Weighted Feature Selection}
To filter out features that are both significant for malware detection and resistant to obfuscation from a high-dimensional pool of candidate features, this study designs the Dynamic Weighted Feature Selection (DWFS) algorithm. DWFS comprehensively evaluates the importance of features on unobfuscated samples and their stability on obfuscated samples, balancing these two aspects through a dynamic weighting mechanism.

\paragraph{Evaluation Of Feature Importance}
On the unobfuscated sample set, a random forest classifier is used to compute the importance of each feature. A classifier is trained using the unobfuscated training data \(X_{unobf}\)and its corresponding labels \(y_{unobf}\). in this paper, the random forest algorithm is employed. Feature importance scores \(I(f_i)\) are extracted from the trained model, where  \(f_i\) represents the i-th feature. Subsequently, the model’s accuracy on the unobfuscated data, denoted as \(acc_{unobf}\), is calculated.

\paragraph{Evaluation Of Feature Stability}
To evaluate the stability of features under different obfuscation techniques, for each obfuscation strategy j (as shown in Table \ref{tb:obfuscate_strategy}), a classifier is trained using the obfuscated training data \(X_{obf_{j}}\) and its corresponding labels \(y_{obj_{j}}\). The model’s accuracy under this obfuscation strategy denoted as \(acc_{obf_{j}}\), is calculated. Feature importance scores \(I_{obf_{j}(f_i)}\) are extracted from the model. Subsequently, the obfuscation impact factor \(\alpha_j\), which represents the degree of impact of obfuscation on model performance, is computed. The calculation of the obfuscation impact factor is shown in Equation \ref{eq:obfuscation_factory}.

\begin{equation}
\label{eq:obfuscation_factory}
\alpha_{j} = \frac{acc_{unobf}-acc_{obf_{j}}}{acc_{unobf}}
\end{equation}

The obfuscation factors \(\alpha_{j}\) for all obfuscation strategies are normalized to ensure \(\Sigma \alpha_j = 1\). Finally, the weighted sum of the importance changes for each feature across all obfuscation strategies is calculated, with the formula shown in Equation \ref{eq:weight_sum}.

\begin{equation}
\label{eq:weight_sum}
\Delta I(f_i) = \Sigma a_j \cdot \left|I(f_i) - I_{obf_{j}}(f_i) \right|
\end{equation}

\paragraph{Dynamic Weighting Mechanism}

Dynamically adjust the weights of importance and stability based on the obfuscation impact factor. A hyperparameter \(\beta\) is set to balance importance and stability. The importance weight is calculated as \(\omega_{1} = 1 - \beta \cdot \bar{a}\), where \(\bar{a}\) is computed as shown in Equation \ref{eq:barA}, with \(m\) representing the number of obfuscation strategies.

\begin{equation}
\label{eq:barA}
\bar{a} = \frac{1}{m} \sum_{j=1}^{m} \alpha_{j}
\end{equation}

\paragraph{Integrated Evaluation}
A composite score is calculated for each feature through a formula that balances its importance and stability, with the calculation method shown in Equation \ref{eq:Integrated_Evluation}. Finally, robust features are \(\theta\) selected based on the composite score, and we set a hyperparameter as the threshold for feature selection. The entire algorithm process is detailed in Algorithm \ref{algorithm:DWFS}.

\begin{equation}
\label{eq:Integrated_Evluation}
S(f_i) = \omega_1 \cdot I(f_i) - \omega_2 \cdot \Delta I f(_i)
\end{equation}

\begin{algorithm}[h]
\label{algorithm:DWFS}
\scriptsize
\SetAlgoLined
\SetKwInOut{Input}{Input}
\SetKwInOut{Output}{Output}
\SetAlgoNlRelativeSize{-1}  
\SetAlgoSkip{0.5em}  
\Input{Unobfuscated data $X_{\text{unobf}}$, labels $y_{\text{unobf}}$, obfuscated data $\{(X_{\text{obf}_j}, y_{\text{obf}_j})\}_{j=1}^{m}$, hyperparameter $\beta$, threshold $\theta$}
\Output{Selected feature subset $F_{\text{selected}}$}
\BlankLine
Train Random Forest on $X_{\text{unobf}}$, $y_{\text{unobf}}$ to obtain importances $I$ and accuracy $\text{acc}_{\text{unobf}}$ 
\tcp*[r]{$I$: importances, $\text{acc}_{\text{unobf}}$: baseline accuracy}

\For{each obfuscation type $j$ from 1 to $m$}{
    Train Random Forest on $X_{\text{obf}_j}$, $y_{\text{obf}_j}$ to obtain importances $I_{\text{obf}_j}$ and accuracy $\text{acc}_{\text{obf}_j}$ \tcp*[r]{$I_{\text{obf}_j}$: importances, $\text{acc}_{\text{obf}_j}$: accuracy}
    
    Compute impact $\alpha_j$ as relative accuracy drop from $\text{acc}_{\text{unobf}}$ to $\text{acc}_{\text{obf}_j}$ \tcp*[r]{$\alpha_j$, obfuscation impact factor}
}
    Compute total impact $\text{sum}_\alpha$ as sum of all $\alpha_j$ 
    
     Normalize each $\alpha_j$ by dividing by $\text{sum}_\alpha$ \tcp*[r]{Normalized weights}
    
    Initialize $\Delta I$ as zero vector matching feature count  \tcp*[r]{$\Delta I$: stability change}
    \For{each $j$ from 1 to $m$}{
        Update $\Delta I$ by adding $\alpha_j$ times absolute difference between $I$ and $I_{\text{obf}_j}$  \tcp*[r]{Weighted stability change}
    }
    
    Compute average impact $\overline{\alpha}$ by averaging all $\alpha_j$ over $m$  \tcp*[r]{$\overline{\alpha}$: mean impact}
    
    Set importance weight $w_1$ by reducing 1 by $\beta$ times $\overline{\alpha}$  \tcp*[r]{$w_1$: importance weight}
    
    Set stability weight $w_2$ as $\beta$ times $\overline{\alpha}$ \; \tcp*[r]{$w_2$: stability weight}
    
    Compute scores $S$ by combining $I$ scaled by $w_1$ and $\Delta I$ scaled by $w_2$, subtracting latter  \tcp*[r]{$S$: feature scores}
    
    Initialize $F_{\text{selected}}$ as empty set  \tcp*[r]{$F_{\text{selected}}$: selected features}
    \For{each feature index $i$}{
        \If{$S_i$ exceeds $\theta$}{
            Add feature $f_i$ to $F_{\text{selected}}$  \tcp*[r]{$f_i$: feature at index $i$}
        }
    }
\Return{$F_{\text{selected}}$}
\caption{Dynamic Weighted Feature Selection (DWFS)}
\end{algorithm}

By comprehensively evaluating the importance and stability of features, and leveraging a dynamic weighting mechanism to adaptively adjust weights, it is possible to filter out a set of features that are both significant and resistant to obfuscation from a vast pool of features. The output features of DWFS can provide high-quality input for subsequent detection models, such as GNNs. thereby enhancing the detection accuracy and robustness of the detection system.

\subsection{Stage 2}
\subsection{Pre-processing}

As shown in Table \ref{tb:FCG_Statistic_Info}, for 8,664 malware samples, the FCG extracted from each APK has, on average, 201,699 nodes and 603,901 edges. If a graph neural network classification method is applied directly to the FCG, the model would need to process a massive amount of node information, resulting in extremely low efficiency and negatively impacting detection accuracy. Therefore, to efficiently capture the malicious behaviors of malware, this paper constructs a method-level Sensitive Behavior Subgraph (SBS). The SBS is a directed graph, which we define as \(G=(V,E)\). Here, \(V\)represents the set of nodes, where each node \(v\in V\) denotes a method, and \(E\) represents the set of edges, where each directed edge \(e=(u,v)\in E\) indicates that method \(u\) calls method \(v\). It is a subgraph of the FCG.

We first extract the FCG from APK files. We use apktool\cite{apkTool} and androguard\cite{androguard} tools for decompilation to obtain the FCG. The SBS is then extracted from the FCG, with the extraction process detailed in Algorithm \ref{algorithm:Get_SBS}.

\subsubsection{SBS Generation}

Previous work relied on sensitive APIs to simplify the FCG\cite{fcg},\cite{sndGCN}. Some of these approaches used BFS or DFS algorithms to traverse all nodes, directly extracting the N-hop neighborhood around sensitive nodes. Through a not particularly rigorous empirical analysis, we found that the aforementioned approaches lead to two issues: one is the inclusion of excessive redundant nodes, which still fails to efficiently capture malicious behaviors; the other is the disconnection of nodes in the graph, preventing the complete preservation of relationships between multiple malicious behaviors. Our method first retains the direct predecessors and successors of sensitive APIs and then optionally extends to an N-hop neighborhood, addressing the above issues. In this paper, sensitive nodes are sourced from the PScout\cite{pscout}

Specifically, the \textbf{direct caller} nodes and \textbf{callee} nodes of sensitive nodes have, on average, K predecessors and L successors. Here, we retain both these predecessor and successor nodes, resulting in a simplified graph with approximately M + M*K + M*L nodes and about M*K + M*L edges, where M is the number of sensitive nodes. When we set N=3, i.e., extending to the 3-hop adjacent nodes of sensitive nodes, calculations show an increase of 16.5\% in the number of nodes and edges, which is still significantly less than the original node and edge counts. The features selected by \ref{subsection:stage1} are stored in the node as node features.

\begin{algorithm}[h]
\label{algorithm:Get_SBS}
\scriptsize
\SetAlgoLined
\SetKwInOut{Input}{Input}
\SetKwInOut{Output}{Output}
\SetAlgoNlRelativeSize{-1}  
\SetAlgoSkip{0.5em}  
\Input{Original function call graph $G = (V, E)$, List of sensitive API nodes $S$, Extension hop count $N$.}
\Output{Simplified function call graph $G' = (V', E')$.}
\BlankLine

\textbf{Initialize:} Set $V' \gets \emptyset$ (set of retained nodes), $E' \gets \emptyset$ (set of retained edges).\\
\For{each sensitive API node $s \in S$}{
    Add $s$ to $V'$.\\
    
    \For{each predecessor $p$ of $s$}{
        Add $p$ to $V'$ and add edge $(p, s)$ to $E'$.\;
    }
    
    \For{each successor $c$ of $s$}{
        Add $c$ to $V'$ and add edge $(s, c)$ to $E'$.\;
    }
    
    \If{$N > 0$}{
        Initialize visited set $\text{visited} \gets \{s\}$ and queue $\text{queue} \gets \text{deque}([(s, 0)])$.\;
        
        \While{$\text{queue}$ is not empty}{
            Dequeue $(u, h)$ from $\text{queue}$.\;
            
            \If{$h \geq N$}{
                \textbf{continue}.\;
            }
            
            \For{each predecessor $p$ of $u$}{
                \If{$p \notin \text{visited}$}{
                    Add $p$ to $V'$ and add edge $(p, u)$ to $E'$.\;
                    Add $p$ to $\text{visited}$ and enqueue $(p, h + 1)$.\;
                }
            }
            
            \For{each successor $c$ of $u$}{
                \If{$c \notin \text{visited}$}{
                    Add $c$ to $V'$ and add edge $(u, c)$ to $E'$.\;
                    Add $c$ to $\text{visited}$ and enqueue $(c, h + 1)$.\;
                }
            }
        }
    }
}

Generate subgraph $G' \gets G.\text{subgraph}(V')$ based on retained nodes and edges.\;

\Return{$G'$} 
\caption{Get Sensitive Behavior Subgraph}
\end{algorithm}

\section{Experimental Evaluation}
\subsection{Datasets and Metrics}

The dataset for this study consists of malicious executable files obtained from AndroZoo\cite{androzoo}. It includes a total of 8,664 malware files (80.558 GB) across eight malware families. Additionally, we utilized the powerful obfuscation tool Obfuscapk\cite{obfuscapk} to obfuscate the malware, employing obfuscation techniques encompassing all the contents listed in Table \ref{tb:obfuscate_strategy}, resulting in 44,940 malware variants. These files were disassembled, and their corresponding features were extracted to obtain the respective SBS.

To comprehensively evaluate the proposed method and assess the effectiveness of the GMC framework, we adopted a series of widely used and representative performance metrics. These metrics include Accuracy, Precision, Recall, and F1 Score.
\subsection{Simplification Effect}
For graph scale simplification, the experimental results are presented in Table \ref{table_FCG_info} and Table \ref{tb:SBS_Statistic_Info}, which provide statistical information on the FCG and SBS for eight malware categories, respectively. Table \ref{table_FCG_info} shows that the FCG is large in scale, with an average node count ranging from 3505.65 to 60254.29, an edge count from 7740.72 to 170902.83, and a total storage space of 203.3 GB. In contrast, Table \ref{tb:SBS_Statistic_Info} indicates that the SBS is significantly reduced, with the average node count dropping to 536 to 6350, the edge count decreasing to 853 to 15408, and the total storage space reduced to only 34.63 GB, a reduction of approximately 83\%. The method I propose, by extracting sensitive behavior subgraphs, not only simplifies the graph structure (reducing the average number of nodes and edges by about 90\%), facilitating analysis, but also significantly improves computational efficiency while lowering storage requirements, making it an efficient and practical solution for large-scale malware detection.

\begin{table*}
\label{tb:FCG_Statistic_Info}
\tiny
\renewcommand{\arraystretch}{1}
\setlength{\tabcolsep}{1.5pt}
\centering
\caption{Function Call Graph Statistic Infomation}
\label{table_FCG_info}
\begin{tabular}{cccccccc} 
\hline
\multicolumn{1}{c|}{\multirow{2}{*}{Class}} & \multicolumn{1}{c|}{\multirow{2}{*}{~Apps}} & \multicolumn{6}{c}{Function Call Graph}                                                           \\ 
\cline{3-8}
\multicolumn{1}{c|}{}                       & \multicolumn{1}{c|}{}                       & \# Graph & Avg \# Nodes & Avg \# Edges & Median \# Nodes~ & Median \# Edges & Storage Space (GB)  \\ 
\hline
adPush                                      & 1500                                        & 1409     & 5809.85      & 16797.31     & 16797.31         & 10386.5         & 8.51                \\
artemis                                     & 1032                                        & 1032     & 3505.65      & 7740.72      & 7740.72          & 2135            & 3.50                \\
openconnection                              & 1494                                        & 1488     & 60254.29     & 170902.83    & 170902.83        & 187081          & 87.50               \\
kuguo                                       & 1500                                        & 1500     & 10414.22     & 34736.42     & 34736.42         & 24961           & 15.35               \\
spyagent                                    & 528                                         & 528      & 16322.40     & 42001.09     & 42001.09         & 45138           & 8.37                \\
dzhtny                                      & 560                                         & 552      & 53210.26     & 167443.68    & 167443.68        & 175712          & 29.18               \\
igexin                                      & 1500                                        & 1500     & 24293.67     & 82755.62     & 82755.62         & 72486           & 35.83               \\
leadbolt                                    & 554                                         & 554      & 27888.68     & 81523.72     & 81523.72         & 77009           & 15.08               \\ 
\hline
Total Statistic                             & 8664                                        & 8563     & 201699.02    & 603901.39    & 198975.50        & 594908.50       & 203.3   
\end{tabular}
\end{table*}

\begin{table*}
\tiny
\renewcommand{\arraystretch}{1}
\setlength{\tabcolsep}{1.5pt}
\centering
\caption{Sensitive Behavior Subgraph Statistic Infomation}
\label{tb:SBS_Statistic_Info}
\begin{tabular}{cccccccc} 
\hline
\multicolumn{1}{c|}{\multirow{2}{*}{Class}} & \multicolumn{1}{c|}{\multirow{2}{*}{~Apps}} & \multicolumn{6}{c}{Function Call Graph}                                                           \\ 
\cline{3-8}
\multicolumn{1}{c|}{}                       & \multicolumn{1}{c|}{}                       & \# Graph & Avg \# Nodes & Avg \# Edges & Median \# Nodes~ & Median \# Edges & Storage Space (GB)  \\ 
\hline
adPush                                      & 1500                                        & 1409     & 536~         & 853~         & 327~             & 385             & 1.14                \\
artemis                                     & 1032                                        & 1032     & 701~         & 1240~        & 136~             & 160             & 1.09                \\
openconnection                              & 1494                                        & 1488     & 6142~        & 13308~       & 6705~            & 14763           & 13.75               \\
kuguo                                       & 1500                                        & 1500     & 1379~        & 2853~        & 957~             & 1673            & 3.12                \\
spyagent                                    & 528                                         & 528      & 1491~        & 3090~        & 1731~            & 3622            & 1.19                \\
dzhtny                                      & 560                                         & 552      & 6350~        & 15408~       & 6516~            & 15462           & 5.29                \\
igexin                                      & 1500                                        & 1500     & 2912~        & 6949~        & 2707~            & 6495            & 6.59                \\
leadbolt                                    & 554                                         & 554      & 2958~        & 5077~        & 3023~            & 5587            & 2.46                \\ 
\hline
Total Statistic                             & 8664                                        & 8563     & 22468        & 48779        & 22102            & 48147           & 34.63              
\end{tabular}
\end{table*}

\subsection{Effectiveness on Classifying General Malware}

The classification results for unobfuscated malware are presented in Table \ref{tb:Unobfuscation_Result}. In terms of overall performance, the GAT model achieves the best results, with an accuracy of 95.56\% and an F1 score of 95.52\%, surpassing GraphSAGE (94.73\% accuracy), GCN (94.47\%), and TAGCN (94.35\%). Across specific families, GAT exhibits outstanding performance on most families, particularly on artemis and kuguo, where its accuracy exceeds 99\%. However, for adpush and spyagent, the F1 scores are slightly lower (91-93\%), suggesting a marginally higher classification difficulty. Notably, on the igexin family, GraphSAGE achieves a higher F1 score (96.21\%) compared to GAT (94.77\%), highlighting differences in model performance on specific tasks. Overall, the features we selected and the methods we proposed demonstrate robust capabilities in the classification of unobfuscated malware.


\begin{table*}
\label{tb:Unobfuscation_Result}
\tiny
\renewcommand{\arraystretch}{1}
\setlength{\tabcolsep}{1pt}
\centering
\caption{Unobfuscation Malware Family Classification Result(\%)}
\begin{tabular}{cc|cccc|cccc|cccc|cccc}
\multicolumn{2}{c|}{}         & \multicolumn{4}{c|}{GAT}      & \multicolumn{4}{c|}{GraphSAGE}    & \multicolumn{4}{c|}{GCN}          & \multicolumn{4}{c}{TAGCN}           \\ 
\hline
\textbf{ID} & \textbf{Family} & Acc   & Pre   & Rec   & F1    & Acc    & Pre    & Rec    & F1     & Acc    & Pre    & Rec    & F1     & Acc    & Pre    & Rec     & F1      \\
1           & adpush          & 97.81 & 93.81 & 92.98 & 93.39 & 97.46~ & 95.27~ & 89.05~ & 92.05~ & 97.14~ & 92.54~ & 89.92~ & 91.21~ & 97.15~ & 91.68~ & 91.28~  & 91.48~  \\
2           & artemis         & 99.75 & 98.32 & 99.63 & 98.97 & 99.53~ & 96.44~ & 99.75~ & 98.07~ & 99.72~ & 98.44~ & 99.27~ & 98.86~ & 99.46~ & 95.68~ & 100.00~ & 97.79~  \\
3           & openconnection  & 99.53 & 94.69 & 98.17 & 96.40 & 99.30~ & 92.12~ & 97.80~ & 94.87~ & 99.28~ & 93.41~ & 95.36~ & 94.37~ & 99.24~ & 92.54~ & 95.91~  & 94.20~  \\
4           & kuguo           & 99.52 & 98.42 & 98.83 & 98.62 & 99.30~ & 97.46~ & 98.59~ & 98.02~ & 99.17~ & 96.55~ & 98.74~ & 97.63~ & 99.40~ & 97.75~ & 98.82~  & 98.28~  \\
5           & spyagent        & 97.80 & 94.57 & 92.76 & 93.66 & 97.33~ & 92.04~ & 92.28~ & 92.16~ & 97.68~ & 93.81~ & 92.96~ & 93.38~ & 97.24~ & 93.42~ & 90.52~  & 91.95~  \\
6           & dzhtny          & 99.36 & 91.99 & 98.38 & 95.08 & 98.93~ & 87.45~ & 97.25~ & 92.09~ & 98.92~ & 87.45~ & 97.30~ & 92.11~ & 99.01~ & 89.42~ & 96.21~  & 92.69~  \\
7           & igexin          & 98.18 & 94.65 & 94.89 & 94.77 & 98.67~ & 97.55~ & 94.91~ & 96.21~ & 98.23~ & 94.56~ & 95.42~ & 94.99~ & 98.39~ & 96.13~ & 94.50~  & 95.31~  \\
8           & leadbolt        & 99.18 & 96.98 & 89.77 & 93.24 & 98.93~ & 92.91~ & 89.62~ & 91.24~ & 98.79~ & 96.94~ & 82.86~ & 89.35~ & 98.80~ & 94.50~ & 86.30~  & 90.21~  \\
9           & Overall         & 95.56 & 95.43 & 95.68 & 95.52 & 94.73~ & 93.91~ & 94.91~ & 94.34~ & 94.47~ & 94.21~ & 93.98~ & 93.99~ & 94.35~ & 93.89~ & 94.19~  & 93.99~  \\
\hline
\end{tabular}
\end{table*}


\subsection{Effectiveness on Classifying Obfuscated Malware}
\begin{table*}
\label{tb:obfuscate_strategy}
\scriptsize
\renewcommand{\arraystretch}{1}
\setlength{\tabcolsep}{2pt}
\centering
\caption{Descriptions of Obfuscators Used in Our Experiments}
\label{tb:FCG_Statistic_Info}
\begin{tabular}{c|c|l} 
\toprule
\multicolumn{2}{c}{Obfuscators}                                                & \multicolumn{1}{c}{Descriptions}                                                                                                      \\ 
\hline
\multirow{4}{*}{\textbf{Trivial}}     & Repackaging                            & \begin{tabular}[c]{@{}l@{}}Unzipping the APK file and re-signing it with a \\different signing certificate.\end{tabular}                                  \\ 
\cline{2-3}
                                      & Disassembling and Reassembling         & \begin{tabular}[c]{@{}l@{}}Disassembling the app using a reverse-engineering tool,\\~By disassembling and reassembling the app.\end{tabular}              \\ 
\cline{2-3}
                                      & Manifest Change                        & \begin{tabular}[c]{@{}l@{}}This transformation changes the manifest by adding \\permissions or adding components’capabilities.\end{tabular}               \\ 
\cline{2-3}
                                      & Alignment~                             & \begin{tabular}[c]{@{}l@{}}This transformation changes the cryptographic hash\\~of an APK file.\end{tabular}                                              \\ 
\hline
\multirow{7}{*}{\textbf{Non-trivial}} & Junk code insertion & Adds code that does not affect the execution of an app.                                                                                               \\ 
\cline{2-3}
                                      & Control-flow manipulation              & \begin{tabular}[c]{@{}l@{}}Changes the methods’control flow graph by adding \\conditions and iterative constructs.\end{tabular}                           \\ 
\cline{2-3}
                                      & Members reordering                     & \begin{tabular}[c]{@{}l@{}}Changes the order of instance variables or \\methods in a classes\textit{.dex }file.~\end{tabular}                             \\ 
\cline{2-3}
                                      & String encryption                      & \begin{tabular}[c]{@{}l@{}}Encrypts the strings in~\textit{classes.dex~}andadds afunction \\that decrypts the encrypted strings at runtime.\end{tabular}  \\ 
\cline{2-3}
                                      & Identifier Renaming                    & \begin{tabular}[c]{@{}l@{}}Renames the instance variables and/or the method \\names in each Java class with randomly generatednames.~\end{tabular}        \\ 
\cline{2-3}
                                      & Class renaming                         & \begin{tabular}[c]{@{}l@{}}Renames the classes and/or the packages in an app \\with randomly generated names.\end{tabular}                                \\ 
\cline{2-3}
                                      & Reflection~                            & \begin{tabular}[c]{@{}l@{}}Transformations convert direct method invocations into \\reflective calls using the Java reflection API.\end{tabular}          \\
\bottomrule
\end{tabular}
\end{table*}


Common obfuscation strategies, as shown in Table \ref{tb:obfuscate_strategy}, are primarily categorized into Trivial Obfuscation and Non-trivial Obfuscation\cite{obfuscationStrategy}. We tested the classification capability of our method under malware obfuscation scenarios. The results are presented in Table \ref{tb:obfuscation_GraphSAGE_Result} and Table \ref{tb:obfuscation_GAT_Result}. Experimental results validate the effectiveness and robustness of combining DWFS with GNNs in classifying obfuscated malware families, achieving an overall performance above 92\%, with near-perfect classification on families such as artemis and igexin. GAT exhibits excellent performance on unobfuscated data (95.56\% accuracy) and maintains a high level on obfuscated data (92.48\% accuracy), confirming the obfuscation resistance of features selected by DWFS. However, obfuscation has a more pronounced impact on certain families (e.g., openconnection).


We also found that the performance gaps between different obfuscation strategies are not significant. This can be attributed to two reasons. First, unobfuscated and obfuscated samples are independently sampled—specifically, a certain number of unobfuscated samples are randomly drawn from each family, followed by independently drawing a certain number of samples from the obfuscated sample pool for each obfuscation strategy, without requiring correspondence to specific unobfuscated samples. This approach prevents direct comparison of feature changes for the same unobfuscated sample across different obfuscation strategies. Second, the features selected by DWFS exhibit such strong robustness against these obfuscation techniques that the variations in feature vectors across different obfuscation strategies are negligible. We plan to address these issues in future improvements.

\begin{table*}
\centering
\tiny
\renewcommand{\arraystretch}{1}
\setlength{\tabcolsep}{1pt}
\caption{Obfuscation Malware Family Classification Result In GraphSAGE Model(\%)}
\label{tb:obfuscation_GraphSAGE_Result}
\begin{tabular}{cc|cccc|cccc|cccc|cccc}
\multicolumn{2}{c|}{}         & \multicolumn{4}{c|}{Encryption}   & \multicolumn{4}{c|}{Rename}       & \multicolumn{4}{c|}{Reflection}   & \multicolumn{4}{c}{Trivial}        \\ 
\hline
\textbf{ID} & \textbf{Family} & Acc    & Pre    & Rec    & F1     & Acc    & Pre    & Rec    & F1     & Acc    & Pre    & Rec    & F1     & Acc    & Pre    & Rec    & F1      \\
1           & adpush          & 90.02~ & 94.79~ & 90.07~ & 92.37~ & 90.03~ & 94.79~ & 90.07~ & 92.37~ & 90.11~ & 94.79~ & 90.07~ & 92.37~ & 90.11~ & 94.79~ & 90.07~ & 92.37~  \\
2           & artemis         & 99.61~ & 96.43~ & 99.61~ & 98.00~ & 99.61~ & 96.43~ & 99.61~ & 98.00~ & 99.61~ & 96.43~ & 99.61~ & 98.00~ & 99.61~ & 96.43~ & 99.61~ & 98.00~  \\
3           & openconnection  & 73.23~ & 84.56~ & 73.33~ & 78.55~ & 73.23~ & 84.56~ & 73.33~ & 78.55~ & 73.23~ & 84.56~ & 73.33~ & 78.55~ & 73.23~ & 84.56~ & 73.33~ & 78.55~  \\
4           & kuguo           & 91.11~ & 93.52~ & 91.11~ & 92.30~ & 91.11~ & 93.52~ & 91.11~ & 92.30~ & 91.11~ & 93.52~ & 91.11~ & 92.30~ & 91.11~ & 93.52~ & 91.11~ & 92.30~  \\
5           & spyagent        & 90.80~ & 93.49~ & 90.80~ & 92.13~ & 90.80~ & 93.49~ & 90.80~ & 92.13~ & 90.80~ & 93.49~ & 90.80~ & 92.13~ & 90.80~ & 93.49~ & 90.80~ & 92.13~  \\
6           & dzhtny          & 97.47~ & 91.41~ & 97.47~ & 94.34~ & 97.47~ & 91.41~ & 97.47~ & 94.34~ & 97.47~ & 91.41~ & 97.47~ & 94.34~ & 97.47~ & 91.41~ & 97.47~ & 94.34~  \\
7           & igexin          & 98.80~ & 96.64~ & 98.80~ & 97.71~ & 98.80~ & 96.64~ & 98.80~ & 97.71~ & 98.80~ & 96.64~ & 98.80~ & 97.71~ & 98.80~ & 96.64~ & 98.80~ & 97.71~  \\
8           & leadbolt        & 96.60~ & 79.01~ & 96.60~ & 86.93~ & 96.60~ & 79.01~ & 96.60~ & 86.93~ & 96.60~ & 79.01~ & 96.60~ & 86.93~ & 96.60~ & 79.01~ & 96.60~ & 86.93~  \\
9           & Overall         & 92.26  & 91.23  & 92.22  & 91.54  & 92.26~ & 91.23~ & 92.22~ & 91.54~ & 92.26~ & 91.23~ & 92.22~ & 91.54~ & 92.26~ & 91.23~ & 92.22~ & 91.54~  \\
\hline
\end{tabular}
\end{table*}

\begin{table*}
\centering
\tiny
\renewcommand{\arraystretch}{1}
\setlength{\tabcolsep}{1pt}
\caption{Obfuscation Malware Family Classification Result In GAT Model(\%)}
\label{tb:obfuscation_GAT_Result}
\begin{tabular}{cc|cccc|cccc|cccc|cccc}
\multicolumn{2}{c|}{}         & \multicolumn{4}{c|}{Encryption}   & \multicolumn{4}{c|}{Rename}       & \multicolumn{4}{c|}{Reflection}   & \multicolumn{4}{c}{Trivial}        \\ 
\hline
\textbf{ID} & \textbf{Family} & Acc    & Pre    & Rec    & F1     & Acc    & Pre    & Rec    & F1     & Acc    & Pre    & Rec    & F1     & Acc    & Pre    & Rec    & F1      \\
1           & adpush          & 94.59~ & 93.64~ & 94.49~ & 94.07~ & 94.51~ & 93.64~ & 94.49~ & 94.07~ & 94.44~ & 93.64~ & 94.49~ & 94.07~ & 94.44~ & 93.64~ & 94.49~ & 94.07~  \\
2           & artemis         & 99.61~ & 97.35~ & 99.61~ & 98.47~ & 99.61~ & 97.35~ & 99.61~ & 98.47~ & 99.61~ & 97.35~ & 99.61~ & 98.47~ & 99.61~ & 97.35~ & 99.61~ & 98.47~  \\
3           & openconnection  & 67.72~ & 76.53~ & 67.72~ & 71.85~ & 67.72~ & 76.53~ & 67.72~ & 71.85~ & 67.72~ & 76.53~ & 67.72~ & 71.85~ & 67.72~ & 76.53~ & 67.72~ & 71.85~  \\
4           & kuguo           & 92.74~ & 95.08~ & 92.74~ & 93.90~ & 92.74~ & 95.08~ & 92.74~ & 93.90~ & 92.74~ & 95.08~ & 92.74~ & 93.90~ & 92.74~ & 95.08~ & 92.74~ & 93.90~  \\
5           & spyagent        & 89.85~ & 96.30~ & 89.85~ & 92.96~ & 89.85~ & 96.30~ & 89.85~ & 92.96~ & 89.85~ & 96.30~ & 89.85~ & 92.96~ & 89.85~ & 96.30~ & 89.85~ & 92.96~  \\
6           & dzhtny          & 99.54~ & 92.01~ & 99.54~ & 95.63~ & 99.54~ & 92.01~ & 99.54~ & 95.63~ & 99.54~ & 92.01~ & 99.54~ & 95.63~ & 99.54~ & 92.01~ & 99.54~ & 95.63~  \\
7           & igexin          & 99.20~ & 98.96~ & 99.20~ & 99.08~ & 99.20~ & 98.96~ & 99.20~ & 99.08~ & 99.20~ & 98.96~ & 99.20~ & 99.08~ & 99.20~ & 98.96~ & 99.20~ & 99.08~  \\
8           & leadbolt        & 96.60~ & 88.28~ & 96.60~ & 92.25~ & 96.60~ & 96.60~ & 88.28~ & 96.60~ & 96.60~ & 88.28~ & 96.60~ & 92.25~ & 96.60~ & 88.28~ & 96.60~ & 92.25~  \\
9           & Overall         & 92.48~ & 92.27~ & 92.47~ & 92.28~ & 92.48~ & 92.27~ & 92.47~ & 92.28~ & 92.48~ & 92.27~ & 92.47~ & 92.28~ & 92.48~ & 92.27~ & 92.47~ & 92.28~  \\
\hline
\end{tabular}
\end{table*}


\section{Conclusion and Future Work}
This paper integrates DWFS with GNNs to construct an efficient and robust malware family detection system. DWFS filters out obfuscation-resistant features from a vast feature pool, while GNNs leverage these features and the program’s function call graph for deep learning. This approach not only improves detection accuracy but also significantly enhances the system’s ability to counter code obfuscation techniques. In the future, we will try to adjust the DWFS strategy, incorporate more sensitive characteristics, combine the feature selection algorithm with different depth learning methods, and further distinguish the confusion method.

%
%
%
%


\end{document}